\documentclass[useAMS,usenatbib,aas_macros,fleqn]{mn2e}

\usepackage{verbatim}

\usepackage{epsfig}
\usepackage{amssymb}
\usepackage{natbib}
\usepackage{aas_macros}
\usepackage{times}

\usepackage{multirow}
\usepackage{xcolor}

\usepackage{graphics}
\usepackage{amsfonts}
\usepackage{amsmath}
\usepackage{layout}
\usepackage{amssymb}
\usepackage{epsfig}

\usepackage{hyperref}
\newcommand{\figu}{Figure~}
\newcommand{\figus}{Figures~}
\newcommand{\eq}{Equation~}
\newcommand{\eqs}{Equations~}
\newcommand{\sect}{Section~}

\newcommand{\kmse}{\rm km\, s^{-1}}

\newcommand{\sis}{$\sigma$}
\newcommand{\sise}{\sigma}

\newcommand{\mbh}{$M_{\rm bh}$}
\newcommand{\mbhe}{M_{\rm bh}}

\newcommand{\ree}{R_e}

\newcommand{\mstar}{$M_{\rm star}$}
\newcommand{\mstare}{M_{\rm star}}
\newcommand{\mbulge}{$M_{\rm bulge}$}
\newcommand{\mbulgee}{M_{\rm bulge}}

\newcommand{\ML}{$M_{\rm star}/L$}

\newcommand{\msune}{M_{\odot}}

\newcommand{\epsie}{\epsilon}

\newcommand{\fvir}{$f_{vir}$}
\newcommand{\fvire}{f_{vir}}

\begin{document}

\voffset=-0.50in

\def\sarc{$^{\prime\prime}\!\!.$}
\def\arcsec{$^{\prime\prime}$}
\def\arcmin{$^{\prime}$}
\def\degr{$^{\circ}$}
\def\seco{$^{\rm s}\!\!.$}
\def\ls{\lower 2pt \hbox{$\;\scriptscriptstyle \buildrel<\over\sim\;$}}
\def\gs{\lower 2pt \hbox{$\;\scriptscriptstyle \buildrel>\over\sim\;$}}

%\title[Probing selections in AGN]{Active galaxies, selection effects, and distinct high virial factors to match the scaling relations of quiescent black holes}

\title[Selection effects and virial factors]{Black hole scaling relations of active and quiescent galaxies: Addressing selection effects and constraining virial factors}

\author[F. Shankar et al.]
{Francesco Shankar$^{1}$\thanks{E-mail:$\;$F.Shankar@soton.ac.uk}, Mariangela Bernardi$^{2}$,
Kayleigh Richardson$^{1}$, Christopher Marsden$^{1}$, \newauthor
Ravi K. Sheth$^{2}$, Viola Allevato$^{3}$, Luca Graziani$^{4}$, Mar Mezcua$^{5,6}$, Federica Ricci$^{7}$,
\newauthor Samantha J. Penny$^{8}$, Fabio La Franca$^{9}$, Fabio Pacucci$^{10}$
% \newauthor
%....$^{4}$
\\
$1$ Department of Physics and Astronomy, University of Southampton, Highfield, SO17 1BJ, UK\\
$2$ Department of Physics and Astronomy, University of Pennsylvania, 209 South 33rd St, Philadelphia, PA 19104\\
$3$ Scuola Normale Superiore, Piazza dei Cavalieri 7, I-56126 Pisa, Italy\\
$4$ Dipartimento di Fisica, Sapienza Universit\`{a} di Roma, Piazzale Aldo Moro 5, I-00185, Roma, Italy\\
$5$ Institute of Space Sciences (ICE, CSIC), Campus UAB, Carrer de Magrans, E-08193 Barcelona, Spain\\
$6$ Institut d'Estudis Espacials de Catalunya (IEEC), Carrer Gran Capit\`{a}, 08034 Barcelona, Spain\\
$7$ Instituto de Astrof\'{\i}sica and Centro de Astroingenier\'{\i}a, Facultad de F\'{\i}sica, Pontificia Universidad Cat\'{o}lica de Chile, Casilla 306, Santiago 22, Chile\\
$8$ Institute of Cosmology and Gravitation, University of Portsmouth, Dennis Sciama Building, Burnaby Road, Portsmouth PO1 3FX, UK\\
$9$ Dipartimento di Matematica e Fisica, Universit\`{a} Roma Tre, via della Vasca Navale 84, I-00146 Roma, Italy\\
$10$ Department of Physics, Yale University, New Haven, CT 06511, USA\\
}
\date{}
\pagerange{\pageref{firstpage}--
\pageref{lastpage}} \pubyear{2018}
\maketitle
\label{firstpage}

\begin{abstract}
Local samples of quiescent galaxies with dynamically measured black hole masses (\mbh) may suffer from an angular resolution-related selection effect, which could bias the observed scaling relations between \mbh\ and host galaxy properties away from the intrinsic relations. In particular, previous work has shown that the observed \mbh-\mstar\ relation is more strongly biased than the \mbh-\sis\ relation.
Local samples of active galactic nuclei (AGN) do not suffer from this selection effect, as in these samples \mbh\ is estimated from megamasers and/or reverberation mapping-based techniques.  With the exception of megamasers, \mbh -estimates in these AGN samples are proportional to a virial coefficient \fvir.  Direct modelling of the broad line region suggests that $\fvire\sim 3.5$.  However, this results in a \mbh-\mstar\ relation for AGN which lies below and is steeper than the one observed for quiescent black hole samples.  A similar though milder trend is seen for the \mbh-\sis\ relation.   Matching the high-mass end of the \mbh-\mstar\ and \mbh-\sis\ relations observed in quiescent samples requires $\fvire\gtrsim 15$ and $\fvire\gtrsim 7$, respectively.  On the other hand, $\fvire\sim 3.5$ yields \mbh-\sis\ and \mbh-\mstar\ relations for AGN which are remarkably consistent with the expected ``intrinsic'' correlations for quiescent samples (i.e., once account has been made of the angular resolution-related selection effect), providing additional evidence that the sample of local quiescent black holes is biased.  We also show that, as is the case for quiescent black holes, the \mbh-\mstar\ scaling relation of AGN is driven by \sis, thus providing additional key constraints to black hole-galaxy co-evolution models.
\end{abstract}


\section{Introduction}
\label{sec|intro}

It has long been accepted that active galactic nuclei (AGN), the central powerhouses of the most energetic galaxies in the Universe, are powered by the release of gravitational energy during the accretion of material onto the supermassive black hole at their centres. The general consensus is that most, if not all, galaxies host a massive black hole and may go through an AGN ``phase''. The latter may be self-regulating both the black hole growth and star formation in the host galaxy via some wind/jet driven feedback mechanisms
\citep[see reviews by, e.g.,][]{ShankarReview,AlexanderHickoxReview}.
Models predict that an AGN and its host may coevolve \citep[e.g.,][]{SilkRees,Granato04}, leading to host characteristics such as galaxy bulge/total stellar mass (\mbulge/\mstar) and/or central stellar velocity dispersion (\sis) being linked to black hole mass (\mbh).  Probing these relations is a primary goal of modern cosmology because its understanding will be a crucial step towards a more complete view of galaxy evolution.

Many groups have quantified scaling relations between supermassive black hole masses and host galaxy properties, in the local Universe, using samples of {\em quiescent} galaxies \citep[see, e.g.,][]{FerrareseFord,KormendyHo,GrahamReview15}. However, \citet{Bernardi07} showed that the \sis-\mstar\ relations defined by quiescent black hole samples differs from that defined by the bulk of the galaxies in the Sloan Digital Sky Survey \citep[][hereafter SDSS]{York00}.  \citet{Remco15} and \citet{Shankar16BH} confirmed that quiescent samples are biased towards dense galaxies.  As a result, it has been suggested that measurements of the \mbh\ scaling relations in these samples may be severely biased \citep[e.g.,][]{Bernardi07,Gultekin09,Bat10,DaiMbhSigma,Shankar16BH}.

In these galaxies, \mbh\ is estimated from the dynamics of stars near the black hole ``sphere of influence'' \citep[e.g.,][]{MerrittFerrareseProceedings}
\begin{equation}
 r_{\rm infl}\equiv G\mbhe/\sigma^2\, .
 \label{eq|rinfl}
\end{equation}
Since $r_{\rm infl}$ is small (order of parsecs), insufficient resolution prevents reliable black hole mass estimates or forces to target only the largest black holes, possibly leading to a selection bias.  \citet{Shankar16BH} used Monte Carlo simulations to show that this resolution-related selection effect has the potential to artificially increase the normalization of the \sis-\mstar\ and \mbh-\sis\ relation by a factor of a few, and the \mbh-\mstar\ by an order of magnitude or more.
Indeed, this selection effect alone can go a long way towards explaining why, to date, pulsar timing arrays have failed to detect any gravitational waves from supermassive black hole mergers \citep{Sesana16}.

To gather more robust constraints on black hole-galaxy coevolution models, it is thus necessary to analyse samples that do not suffer from this resolution-related selection effect.
%If these other samples follow the same scaling relations with host galaxy properties defined by the bulk of the population, then this will strengthen the hypothesis that quiescent samples are biased \citet{Shankar16BH}.
In active galaxies, \mbh\ can be estimated spectroscopically from the kinematics of the broad line region or gas orbiting around the very inner regions around the central black hole \citep[e.g.,][]{FerrareseFord}.  However, except for masers for which black hole masses can be inferred from rotation curves \citep[e.g.,][]{FerrareseFord}, for the vast majority of (Type 1) AGN, black hole masses are derived from the (presumed) virial motions of the broad line region gas cloud orbiting in the vicinity of the central compact object:
\begin{equation}
    \mbhe=\fvire \frac{r\left(\Delta V \right)^2}{G}.
    \label{eq|fvir}
\end{equation}
In \eq\ref{eq|fvir}, $r$ is the radius of the broad line region (BLR), which is derived from reverberation mapping \citep[e.g.,][]{Blandford82,Peterson93}, or reverberation-based methods that make use of the radius-luminosity relation \citep[e.g.,][]{Bentz06}. The characteristic velocity $\Delta V$ is derived from the width of the emission lines (a common one is H$\beta$), and $G$ is the gravitational constant. As motions in the BLR are not perfectly Keplerian, a parameter \fvir\ is included in \eq\ref{eq|fvir} to account for the uncertainties in kinematics, geometry and inclination of the clouds \citep[e.g.,][and references therein]{HoKim14}.
Systematically different values of \fvir\ can be found if the virial masses (based on reverberation mapping campaigns) are computed using the full width at half maximum or dispersion of the emission line \citep[e.g.,][]{Onken04,Collin06}. In what follows, we will always refer to \fvir\ as the virial factor calibrated on the line dispersion (i.e., second moment of the line profile) of H$\beta$ \citep[e.g.,][]{Peterson04}.

A reasonable guess for \fvir\ is obtained by matching the AGN black hole scaling relations to those of inactive black holes. Typical values for line dispersion-based \fvir\ derived from the match to the \mbh-\sis\ relation of quiescent black holes are in the range $\fvire\sim 4-5$ \citep[e.g.,][and references therein]{Onken04,Woo10,Grier13,Batiste17}. Although lower values for \fvir\ have also been claimed \citep[][]{Graham11}, \citet{HoKim14} invoke larger values, $\fvire\sim 6-10$, when comparing AGN in classical bulges to the quiescent sample of \citet{KormendyHo}.  In cases in which the BLR structure and dynamics have been modelled directly, the virial coefficient can be constrained directly from the data.
Following the geometric and dynamic modelling put forward by \citet{Pancoast14}, \citet{Grier17} have fitted the line dispersion of the H$\beta$ emission-line spectra of four sources finding a mean $\fvire\sim 3.5$ ($\log \fvire=0.54\pm 0.17$).  \citet{Williams18} have more recently confirmed an average value of $\fvire\sim 3.7$ ($\log \fvire=0.57\pm 0.19$) when fitting the H$\beta$ line dispersion of another seven AGN. The latter values are systematically lower than those found by, e.g, \citet[][]{HoKim14}. Clearly, to date, there is no consensus on the best value of \fvir\ for weighing black holes in AGN.

To use AGN as reliable probes of black hole-galaxy scaling relations, more secure determinations of \fvir\ are required.  The main aim of this work is to take a step in this direction.  To this end, we analyse scaling relations in a variety of local AGN samples, so as to probe their level of bias with respect to a larger sample of local galaxies from the SDSS.  We briefly describe how we homogenize the samples in \sect\ref{sec|data}, and provide full details in Appendix~\ref{Appendix}.  Section~\ref{sec|results} presents scaling relations of the properties of the galaxies which host AGN -- to address the question of whether these are a biased subset -- before considering scaling relations with \mbh\ and their implications for \fvir.  It should be noted in fact that, since resolution-related selection effects have a stronger impact on the quiescent \mbh-\mstar\ relation rather than on the \mbh-\sis\ one \citep{Shankar16BH}, if such selection effects are present, systematically different mean values of \fvir\ should be derived when calibrating AGN samples against the \mbh-\sis\ or the \mbh-\mstar\ relations of local inactive black hole samples. In this work we show that such a systematic offset in \fvir\ is indeed present, but the offset can be removed by accounting for the resolution-related selection effect.

%Moreover, the Monte Carlo simulations performed by \citet{Shankar16BH} pointed out that the observational selection effects imposed by the small black hole sphere of influence, mirrored in the offsets in the \sis-\mstar, propagate more strongly in the \mbh-\mstar, which ends up with an observed higher normalization than the one characterizing the observed \mbh-\sis\ relation.

Throughout we adopt $h=0.7$, $\Omega_m=0.3$, $\Omega_{\Lambda}=0.7$ for the SDSS sample. All AGN samples considered here have black holes and host galaxy properties derived assuming identical or very similar cosmological parameters\footnote{The largest offset is found for the \citet{HoKim14} sample who adopt $h=0.705$ and $\Omega_m=0.27$ for calibrating distances, which anyway induce only $\sim 0.5\%$ difference in distances at the redshifts of interest here.}.

%\section{Homogenizing different data sets in stellar mass and black hole mass}
\section{Homogenizing different data sets}
\label{sec|data}

In this work we consider a number of serendipitous local AGN samples.  These were collected at different flux limits and with different host galaxy morphologies.  Therefore, it is important to homogenize them so that their \mstar\ and \mbh\ values can be meaningfully compared. We clarify that throughout this paper \mstar\ always refers to the \emph{total} stellar mass of a galaxy, while \mbulge\ refers to its bulge stellar mass.

For our study, we correct all total stellar mass estimates so they correspond to \ML\ from \citet{Bell03} and a Chabrier \citep{Chabrier03} IMF, since this was the choice made by \citet{Shankar16BH} in their study of the {\it intrinsic} (i.e., unbiased) black hole scaling relations.  We make use of velocity dispersions\footnote{Unless otherwise noted, throughout this work \sis\ always refers to the velocity dispersion within one effective radius $R_e$, $\sigma_e \equiv\sigma(R_e)$.} $\sigma=\sigma_e=\sigma(R_e)$ measured within the galaxy optical effective radius $R_e$. For the quiescent and some active black hole samples, we correct velocity dispersions measured within an aperture of radius $R$ to the value they are expected to have within an aperture of one effective radius using the S\'{e}rsic index-dependent correction $\sigma(R_e) \propto (R/R_e)^{\gamma(n_{\rm Ser})}$ \citep{Bernardi17IMF}. Finally, if not otherwise specified (as for the sample from \citealt{Mezcua18}), wherever relevant we assume $\fvire=3.5$, following the Lick Monitoring Project \citep{Walsh09} and the latest results of \citet{Grier17} and \citet{Williams18} when adopting the H$\beta$ line dispersion.  The Appendix provides details of these corrections for the different samples we use.
A list of the datasets considered in this work is given in Table \ref{DatasetRefTable}.

\begin{table}
\begin{tabular}{|l|l|l|l|}
\hline
 Data acronym & Method     & Reference                                                                       \\ \hline
\hline
 &  Active
\\ \hline
 H\&K    & Reverberation        & \cite{HoKim14} \\
 MN\&M   & Single Epoch        & \cite{Mezcua18} \\
 VdB     & Single Epoch & \cite{Remco16} \\
  VdB     & Masers & \cite{Remco16} \\
 G+      & Masers        & \cite{greene2016megamaser} \\
 B+      & Single Epoch        & \cite{busch2014low} \\
 R\&V    & Single Epoch        & \cite{ReinesVolonteri15} \\
 B\&MN   & Single Epoch        & \cite{Bentz18}                                        \\ \hline
  &  Quiescent
\\ \hline
 Sa+     & Dynamical        & \cite{SavorgnanGraham16}    \\
 K\&H    & Dynamical        & \cite{KormendyHo}             \\                      \hline
\end{tabular}
\caption{\label{DatasetRefTable} List of the datasets used throughout this paper.}
\end{table}

\begin{figure*}
    \center{\includegraphics[width=\textwidth]{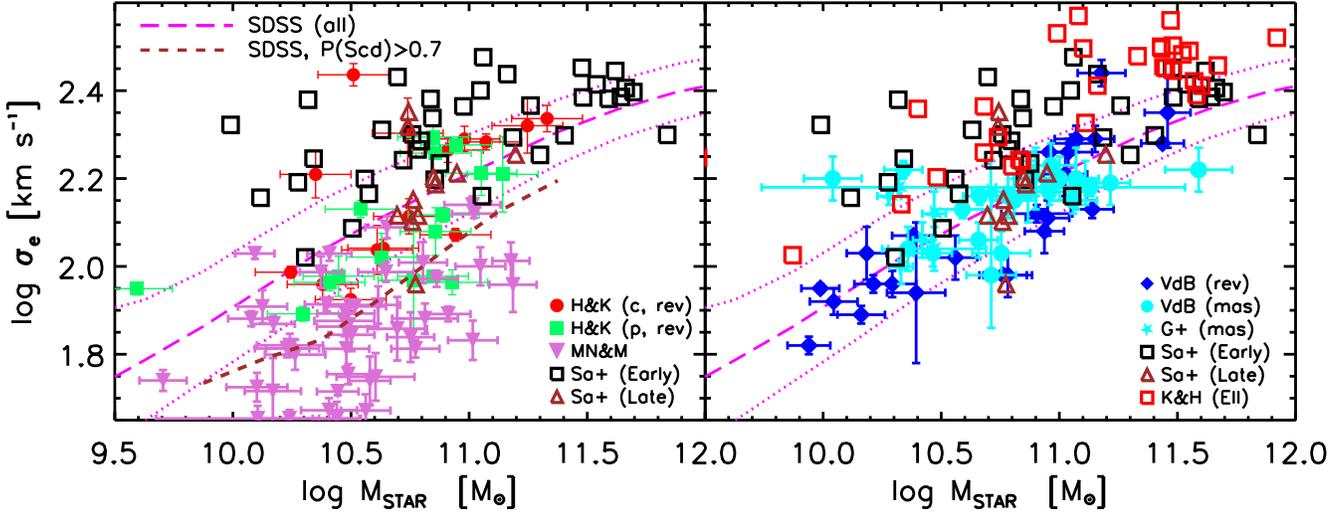}
    \caption{\emph{Left}: Correlation between velocity dispersion and (total) stellar mass in SDSS for all galaxies (long-dashed magenta line; the dotted magenta lines show the region enclosing 68\% of the objects at each \mstar) and also for later-type galaxies with a high probability $P$(Scd) $>0.7$ of being classified as late spirals (brown dashed line). The data are the classical and pseudobulges from \citet[][red circles and green squares]{HoKim14}, and the low-mass Seyferts from \citet[][orchid upside down triangles]{Mezcua18}. \emph{Right}: Same correlation between velocity dispersion and (total) stellar mass in SDSS for all galaxies, compared to the reverberation-based and masers from \citet[][blue diamonds and cyan circles]{Remco16}, and
    the megamasers from \citet[][cyan stars]{greene2016megamaser}. As in the left panel, open black squares and brown triangles are from \citet{SavorgnanGraham16}, while the open red squares are the elliptical quiescent supermassive black hole sample from \citet{KormendyHo}. In contrast to the quiescent galaxies, AGN galaxies seem to follow very similar scaling relations to SDSS galaxies. For simplicity, in this and in all subsequent figures, masers are labelled as ``mas'', while all other AGN samples with black hole masses based on direct reverberation mapping or reverberation-based (i.e., single-epoch/virial) techniques, are labelled as ``rev''.
    \label{fig|FigureSigmaMstar}}}
\end{figure*}

\section{Results}
\label{sec|results}

\subsection{The \sis-\mstar\ relation of active galaxies}
\label{subsec|SigmaMstar}

The first step of our analysis is to compare the \sis-\mstar\ relation of the different AGN samples among themselves and with that of a much larger sample of
galaxies from the SDSS.
The purpose of this check is twofold. First we ensure that, at least for the AGN samples with available (total) stellar mass and velocity dispersion measurements, the homogenizing corrections mentioned in \sect\ref{sec|data}, and expanded in Appendix~\ref{Appendix}, provide consistent results. Second, we probe by how much, if at all, active host galaxies differ in terms of stellar mass and/or velocity dispersion from the general population of SDSS galaxies. In fact, as mentioned in \sect\ref{sec|intro}, \citet{Bernardi07} and \citet{Shankar16BH} showed that local inactive galaxies with dynamical black hole mass measurements (open black and red squares and brown triangles in \figu\ref{fig|FigureSigmaMstar}) tend to have, on average, larger velocity dispersions compared to counterpart galaxies in the SDSS. This behaviour can be in large part explained by the selection effect imposed by limited telescope resolution power, which forces to preferentially target more massive black holes, predominantly hosted in galaxies with larger velocity dispersions \citep{Shankar16BH,Shankar17BH,Barausse17}.

\figu\ref{fig|FigureSigmaMstar} shows\footnote{For simplicity, in this and in all subsequent Figures, to distinguish them from masers (labelled as ``mas''), we label as ``rev'' all AGN samples in which black hole masses are based on direct reverberation mapping or reverberation-based (i.e., single-epoch virial) techniques.} that reassuringly, after applying the proper homogenizing corrections in stellar mass and velocity dispersion,
almost all the classical and pseudobulges from \citet[][filled red circles and green squares, respectively]{HoKim14},
the megamasers from \citet[][filled cyan stars]{greene2016megamaser}, and the active galaxies in the
\citet[][blue diamonds and cyan circles for reverberation-based and masers AGN, respectively]{Remco16} sample, fall within the scatter
(see also \citealt[][their \figu4,]{Grier13} for a similar result) of the SDSS velocity dispersion-stellar mass relation (magenta long-dashed and dotted lines).
For the SDSS galaxies we adopt the latest rendition of the \citet{Meert15} galaxy sample,
with light profiles extracted from S\'{e}rsic+exponential models and mass-to-light ratios from \citet{Mendel13}. We increase all stellar masses by an average 0.05 dex (see \figu A2 in \citealt{Bernardi17}) to account for the (small) difference at high stellar masses with the \citet{Bell03SEDs} mass-to-light ratios used by \citet{Shankar16BH}
and adopted as a reference in this work. Some of the lower mass pseudobulges from \citet[][green squares]{HoKim14} and all the low-mass Seyferts
from \citet[][orchid upside down triangles]{Mezcua18} tend to have, on average, somewhat lower velocity dispersions at fixed stellar mass by $\lesssim 0.05-0.1$ dex, better lining up, as expected, with the SDSS \sis-\mstar\ relation of later-type galaxies (brown dashed line). The latter was calibrated out of the subsample of SDSS galaxies with a high probability $P$(Scd) $>0.7$ of being classified as late spirals according to the Bayesian automated classification of \citet{Huertas11}.

\begin{figure*}
    \center{\includegraphics[width=\textwidth]{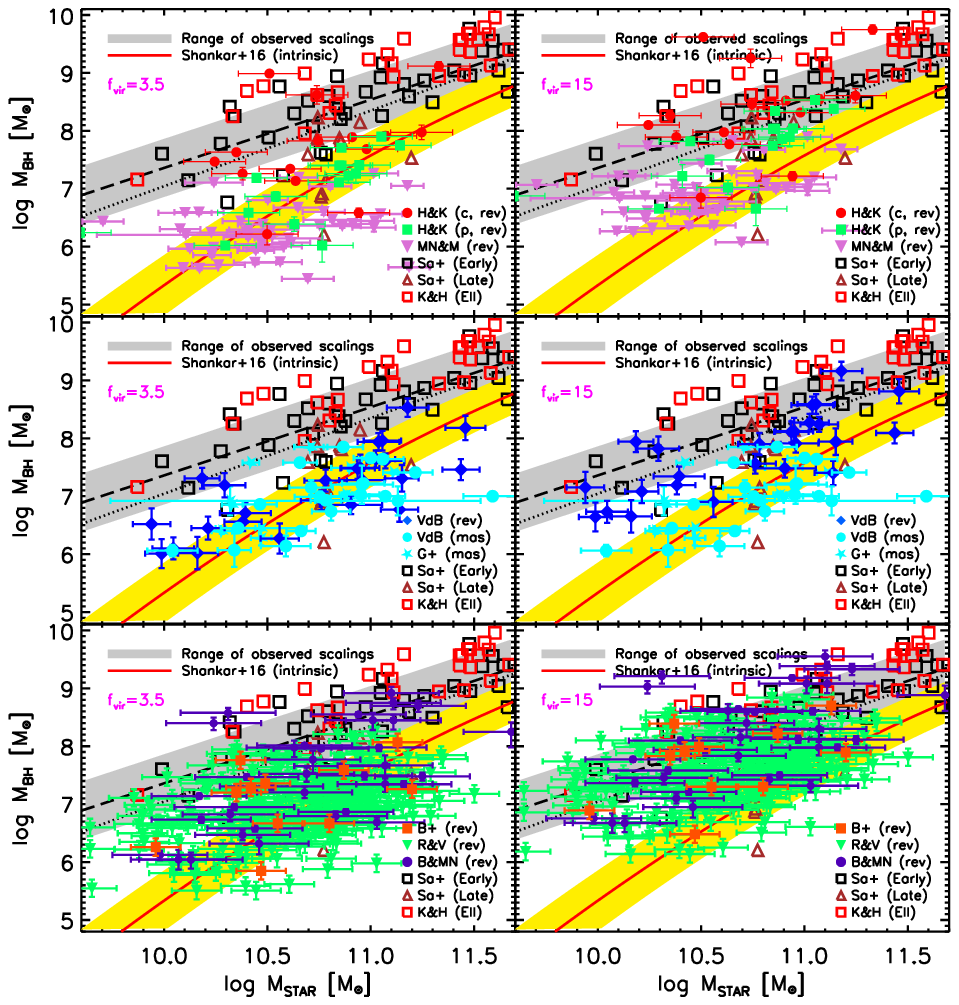}
    \caption{\emph{Left panels}: Correlations between \mbh-\mstar\ in quiescent galaxies (the black dotted line shows the fit to the all sample; black dashed line and grey band show the fit and rms-scatter for the early-type quiescent galaxies subsample, respectively) with data from
    \citet[][open black squares and brown triangles]{SavorgnanGraham16} and the ellipticals from
    \citet[][open red squares]{KormendyHo}, and the unbiased \mbh-\mstar\ relation predicted by \citet[][solid red line and yellow band]{Shankar16BH}.
    The correlations are compared with a variety of samples of active galaxies for which a mean virial factor $\fvire=3.5$ was used. The samples are from
    \citet[][red circles and green squares for classical and pseudobulges, respectively]{HoKim14},
    \citet[][orchid upside down triangles]{Mezcua18}, in the top panel, \citet[][blue diamonds and cyan circles]{Remco16},
    \citet[][cyan stars]{greene2016megamaser}, in the middle panel, and \citet[][red squares]{busch2014low}, \citet[green upside down triangles][]{ReinesVolonteri15},
    and \citet[][purple circles]{Bentz18}, in the bottom panel. \emph{Right panels}: Same format as left panels but adopting a mean virial factor of $\fvire=15$ for all virial-based active black hole mass measurements. When adopting the mean virial factor of $\fvire=3.5$ as constrained by \citet{Grier17}, most of the active galaxies tend to better line up with the \citet{Shankar16BH} ``intrinsic/unbiased'' \mbh-\mstar\ relation (left panels), while substantially larger virial factors are required to match the \mbh-\mstar\ relation of quiescent galaxies (right panels).
    \label{fig|FigureMbhMstarAGN}}}
\end{figure*}

In the right panel of \figu\ref{fig|FigureSigmaMstar} we also include the black hole sample\footnote{We slightly correct the stellar masses in the \citet{KormendyHo} sample to match the $K$-band mass-to-light ratios to convert from dynamical to stellar masses and the same IMF (see text after \eq\ref{eq|MLfromKH}). Velocity dispersions are also corrected from $0.5\ree$ to $\ree$ following \citet{Cappellari06}. We note that such corrections are relatively small and, if neglected, would further exacerbate the tension between the \citet{KormendyHo} sample and the SDSS and \citet{Remco16} galaxies.} of elliptical galaxies from
\citet[][red open squares]{KormendyHo}. First off, it is relevant to note that the \citet{KormendyHo} ellipticals broadly overlap with the early-types from the \citet[][open black squares]{SavorgnanGraham16} black hole sample and lie substantially above the mean \sis-\mstar\ relation of SDSS galaxies. \citet{Shankar16BH} also showed that other samples of black holes in inactive galaxies \citep[][]{Beifiori12,McConn13,Laesker14} lie systematically above the SDSS \sis-\mstar\ relation, further supporting the view that such a mismatch is not a result of different choices in stellar mass calibrations and/or apertures, but an underlying selection effect affecting the sample of black holes in local inactive galaxies with dynamical mass measurement. It is particularly meaningful to compare the \citet{KormendyHo} ellipticals with the \citet{Remco16} sample. Both samples adopt exactly the same mass-to-light ratio and span a similar range in stellar mass $10\lesssim \log \mstare/\msune\lesssim 11.5$, yet the \citet{KormendyHo} data points are placed at significantly larger velocity dispersions. Whilst some of the offset between the two samples can be ascribed to simple morphological segregation, being the \citet{Remco16} AGN sample dominated by Seyfert-like galaxies which tend to be characterised by lower velocity dispersions (see left panel), still this effect is on average relatively small to account for the full offset. Within the SDSS later-type galaxies, in fact, only Scd galaxies tend to show a moderate offset of $\sim 0.1$ dex in velocity dispersion at fixed stellar mass with respect to the full SDSS galaxy population (left panel). On the other hand, Sab galaxies, which are much more common among (especially Type 1) Seyferts \citep[e.g.,][]{Chen17review}, share, we checked, a very similar \sis-\mstar\ relation to the general population.

All in all, from \figu\ref{fig|FigureSigmaMstar} we conclude that the samples of AGN considered in this work are \emph{not} biased in terms of their velocity dispersions and stellar masses when compared to the SDSS galaxies. In this respect, being more faithful tracers of the local galaxy population, AGN samples appear as a more robust ``training set'' to probe the underlying scaling relations between black holes and their host galaxies. The systematic discrepancies, which we will discuss below, between the scaling relations of active and quiescent galaxies in view of \figu\ref{fig|FigureSigmaMstar} should be mostly ascribed to selections rather than to physical effects. In the framework of the selection effects discussed above, if galaxies are offset in the SDSS \sis-\mstar\ relation, they should show a higher degree of bias in the \mbh-\mstar\ relation rather than in the \mbh-\sis\ relation. As in fact evidenced from the Monte Carlo simulations by \citep[][]{Shankar16BH},
selecting sources with larger \sis, above the mean \sis-\mstar\ correlation, would naturally correspond to substantially more massive black holes,
due to the steepness and tightness of the underlying \mbh-\sis\ relation.
At fixed velocity dispersion the effect of the bias would be less prominent as
correlations with the variable on which the selection was made (in this case \sis) will naturally be less biased.
However, the AGN samples considered in this work, which share a similar \sis-\mstar\ relation to SDSS galaxies,
should present self-consistent black hole scaling relations in terms of velocity dispersion and stellar mass. We will show below that this is indeed the case.

\citet{Bentz18} have recently pointed out that the presence of bars in many local galaxies could induce a bias in the SDSS galaxy sample itself.
Indeed the number of bars in the SDSS galaxies is significant \citep[e.g.,][]{Consolandi16}.
Bars, as also pointed out by \citet{Shankar16BH}, may increase velocity dispersions at fixed stellar mass (see also \citealt{Graham11} and \citealt{Batiste17}).
Nevertheless, even if such an effect is present, this would imply an intrinsic/unbiased SDSS \sis-\mstar\ relation lower in normalization than what plotted in \figu\ref{fig|FigureSigmaMstar}, which would exacerbate the tension with the local sample of black holes in quiescent galaxies. The good match with the AGN hosts tends to either disfavour the presence of a strong bar-induced bias in SDSS galaxies, or points to a similar fraction of barred galaxies among the different samples.

\subsection{The \mbh-\mstar\ relation of local active galaxies}
\label{subsec|localMbhMstar}

In each panel of \figu\ref{fig|FigureMbhMstarAGN} we report the \mbh-\mstar\ relation from \citet{Shankar16BH} of quiescent black holes as extracted from the \cite{SavorgnanGraham16} sample with updated black hole masses from \citet{KormendyHo}, and (total) stellar masses from full bulge-disc decompositions and 3.6 $\mu$m mass-to-light ratios $(M/M_\odot)/(L/L_\odot)=0.6$ from \citet[][]{Meidt14}.
The black open squares and brown triangles represent, as in \figu\ref{fig|FigureSigmaMstar}, early and late type galaxies, respectively. As discussed in \citet{Shankar16BH}, we retain from the original \cite{SavorgnanGraham16} sample only the galaxies with secure black hole mass measurements and remove those sources classified as ongoing mergers, limiting the final sample to 48 galaxies out of which 37 are early-type galaxies (ellipticals or lenticulars). The black long-dashed and dotted lines are, respectively, the linear fits to the \cite{SavorgnanGraham16} early type and full sample, respectively
\begin{equation}
\log \frac{\mbhe}{\msune} =
 8.54 + 1.18\,\log \left(\frac{\mstare}{10^{11}\msune}\right)
\label{eq|ObservedMbhMstar}
\end{equation}
and
\begin{equation}
\log \frac{\mbhe}{\msune} =
 8.35 + 1.31\,\log \left(\frac{\mstare}{10^{11}\msune}\right)\, ,
\label{eq|ObservedMbhMstarAll}
\end{equation}
while the grey area (corresponding to a scatter of $0.5$ dex) broadly brackets the uncertainty region around the observed \mbh-\mstar\ observed in local early-type quiescent galaxies. For completeness, we also plot the elliptical galaxies from the original \citet[][open red squares]{KormendyHo} sample which all lie within, if not even above, the grey band. We indeed verified that all the local quiescent black hole samples considered by \citet{Shankar16BH} fall within the grey band.

The solid red line is instead the intrinsic/unbiased (or debiased) black hole mass versus total stellar mass (\mbh-\mstar) relation proposed by \citet{Shankar16BH}
\begin{multline}
\log \frac{\mbhe}{\msune} =
 7.574 + 1.946\,\log \left(\frac{\mstare}{10^{11}\msune}\right)- 0.306\\
\times\left[\log\left(\frac{\mstare}{10^{11}\msune}\right)\right]^2
 - 0.011\,\left[\log\left(\frac{\mstare}{10^{11}\msune}\right)\right]^3\, ,
 \label{eq|IntrinsicMbhMstar}
\end{multline}
with a mass-dependent intrinsic scatter (yellow region) given by
\begin{equation}
\Delta \log \frac{\mbhe}{\msune} =
 0.32 - 0.1\times \log \left(\frac{\mstare}{10^{12}\msune}\right)\, .
 \label{eq|IntrinsicScatterMbhMstar}
\end{equation}
(\eq\ref{eq|IntrinsicMbhMstar} is applicable to galaxies with stellar mass above $\log \mstare/\msune>10$). The \citet{Shankar16BH} relation (labelled as ``intrinsic'') is significantly lower in normalization and steeper than the relation that would be inferred by directly fitting the \cite{SavorgnanGraham16} data (long-dashed and dotted lines).
This different shape in the intrinsic \mbh-\mstar\ relation mainly reflects the curvature in the \sis-\mstar\ relation (\figu\ref{fig|FigureSigmaMstar}),
and the fact that the black hole mass is tightly related to velocity dispersion.
We note that, strictly speaking, the unbiased relation put forward by \citet{Shankar16BH} has been calibrated against the SDSS early-type galaxies, and thus it should be preferentially compared to the black long-dashed line. Nevertheless, late-type galaxies become progressively subdominant at stellar masses $\log \mstare/\msune \gtrsim 10.5$ \citep[e.g.,][]{Bernardi13}, and if included they would tend, if anything, to decrease the mean velocity dispersions by $\lesssim 0.05$ dex at the stellar masses considered in this work. In turn, the unbiased \mbh-\mstar\ relation would then
be proportionally lower at fixed host stellar mass, thus further increasing the mismatch with the scaling relation of quiescent galaxies.
In what follows we will thus continue to safely consider the yellow band as the region encompassing the intrinsic \mbh-\mstar\ relation of the full local central black hole population hosted in relatively massive galaxies.

\begin{figure*}
    \center{\includegraphics[width=\textwidth]{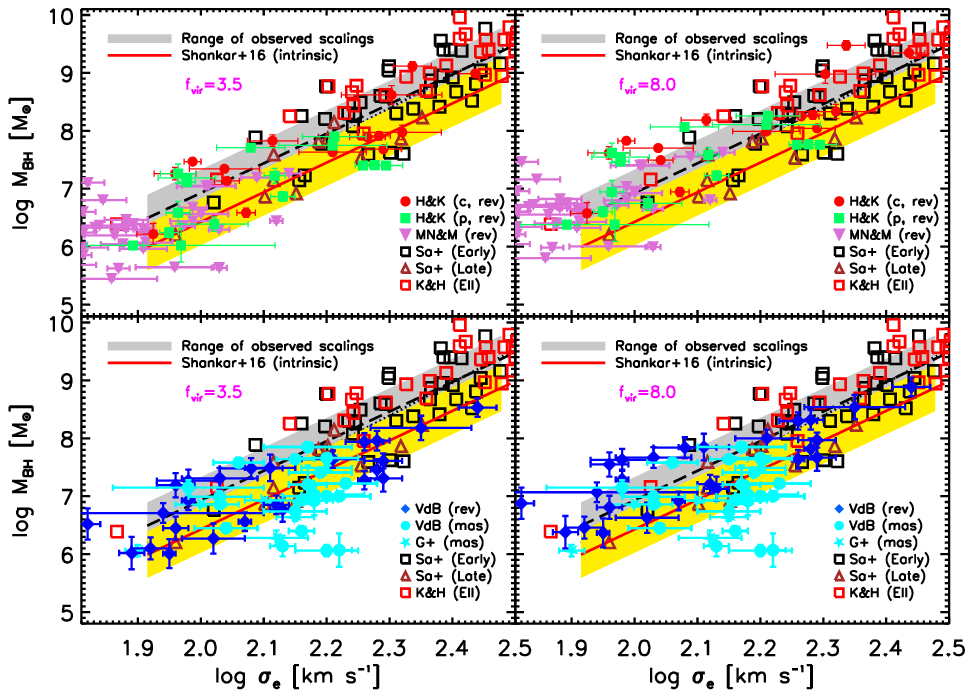}
    \caption{Same format as \figu2 but for the correlations between \mbh-\sis\ in the \citet{SavorgnanGraham16} quiescent galaxies (black lines and grey band)
    and the ``intrinsic'' (unbiased) \mbh-\sis\ relation predicted by \citet[][solid red line and yellow band]{Shankar16BH}.
    When adopting the mean virial factor of $\fvire=3.5$ as constrained by \citet{Grier17}, most of the data tend to better line up with the \citet{Shankar16BH} ``intrinsic/unbiased'' \mbh-\sis\ relation (left panels), while substantially larger virial factors are required to match the \mbh-\sis\ relation of quiescent galaxies (right panels). More quantitative estimates of the appropriate virial scaling factors are given in \figu4.
    \label{fig|FigureMbhSigmaAGN}}}
\end{figure*}

It should be noted that in \figu\ref{fig|FigureMbhMstarAGN} we include all the AGN data sets detailed in Appendix~\ref{Appendix}.
In the upper panels we include the classical and pseudobulges reverberation-based black hole masses from
\citet[][filled red circles and green squares, respectively]{HoKim14} and the low-mass Seyferts from \citet[][orchid upside down triangles]{Mezcua18}, in the middle panels the the megamasers from \citet[][filled cyan stars]{greene2016megamaser}, and the virial-based (i.e., single epoch) and masers from
\citet[][blue diamonds and cyan circles for reverberation-based and masers AGN, respectively]{Remco16}, and in the lower panels the virial-based black holes in low-luminosity quasars from \citet[][red filled squares]{busch2014low},
the virial-based AGN from \citet[][purple circles]{Bentz18}, and the SDSS virial-based AGN from \citet[][green upside down triangles]{ReinesVolonteri15}.

It is clear from the left panels of \figu\ref{fig|FigureMbhMstarAGN} that when adopting a uniform virial factor of $\fvire=3.5$, as independently calibrated by \citet{Grier17} and \citet{Williams18} via direct broad line region modelling, the vast majority of the data tend to lie on average within the yellow band of \citet{Shankar16BH}. However, we stress that the latter is \emph{not} a fit to the AGN samples. Indeed, some of the data sets present larger scatter than what implied by the yellow band, as evident for example in the \citet{Mezcua18} and \citet{ReinesVolonteri15} sources. The \citet{ReinesVolonteri15} and \citet{Remco16} samples also tend to be better fitted by a linear relation between black hole and stellar mass, somewhat flatter than the nearly quadratic one suggested by \eq\ref{eq|IntrinsicMbhMstar}. On the other hand, the \citet{Bentz18} sources are consistent with a similarly steep relation. It is worth noticing that some among the most massive ($\mstare \gtrsim 10^{11}\, \msune$) sources in \citet{Bentz18} lie above \eq\ref{eq|IntrinsicMbhMstar}, lining up with the observed \mbh-\mstar\ relation, though however the sample taken as a whole still points to an average lower \mbh-\mstar\ scaling relation. We note that \citet{Bentz18}'s choice of limiting the S\'{e}rsic index to $n_{\rm Ser}\le4$ could lead to underestimate the luminosities of their most massive galaxies which are usually characterized in SDSS by progressively higher S\'{e}rsic indices \citep[e.g.,][]{Bernardi12}. In this respect, it is worth stressing that all the galaxies in the \citet{busch2014low} sample, which is a mixture of early- and late-type galaxies, all lie significantly below the observed scaling relations.

An average virial factor of $\fvire=3.5$ seems insufficient to reconcile active black holes to the scaling relations observed for quiescent black holes (black lines). As shown in the right panels of \figu\ref{fig|FigureMbhMstarAGN}, we would in fact need to increase the virial factor by more than an order of magnitude, e.g. $\fvire\gtrsim 15$, in order to be roughly consistent with the observed \mbh-\mstar\ relations (black dotted and long-dashed lines), at least around $\log \mstare/\msune\gtrsim 11$. At lower stellar masses the AGN data would require even higher virial factors due to the stronger steepness of their \mbh-\mstar\ relation.
A more quantitative analysis of the appropriate virial factors needed to match the scaling relations of quiescent black holes will be presented in \sect\ref{subsec|fvir}.

It is worth noticing that when increasing the virial factor only the virial-based black hole masses are proportionally varied but not the ones based on independent maser measurements. This creates a systematic discrepancy which is evident in the middle, right panel of \figu\ref{fig|FigureMbhMstarAGN} between the masers by \citet[][cyan stars]{greene2016megamaser} and \citet[][cyan circles]{Remco16} with the virial-based black holes from the same
\citet[][blue diamonds]{Remco16} sample. This offset is not present when $\fvire=3.5$ is adopted (middle left panel),
thus lending further support to the presence of a bias in inactive black holes.

\begin{figure*}
    \center{\includegraphics[width=\textwidth]{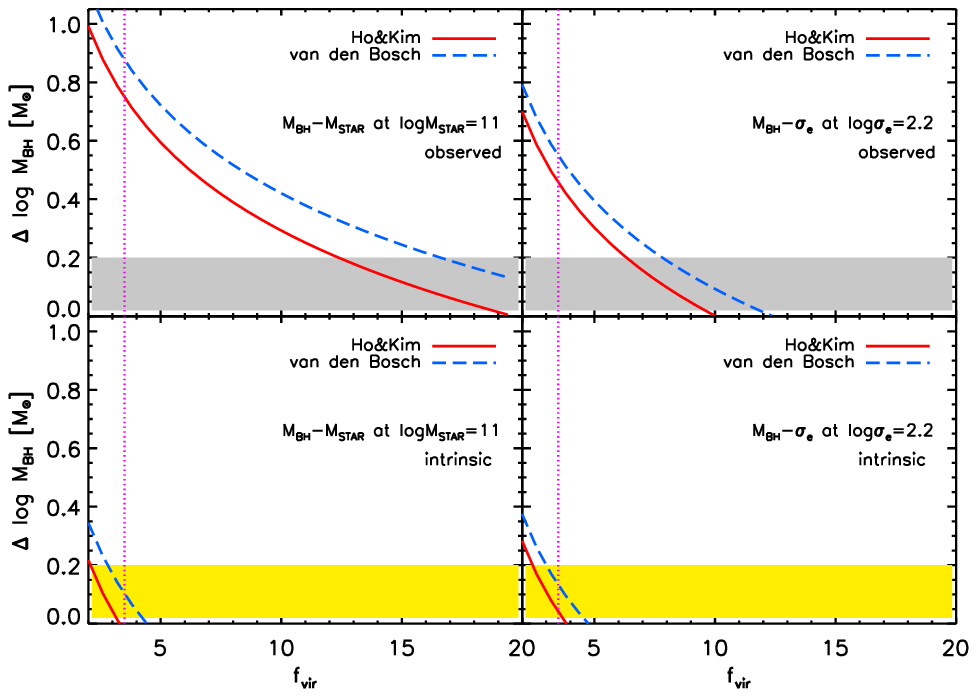}
    \caption{\emph{Left}: Displacement in $\log \mbhe$ between the $\log \mbhe-\log \mstare$ relation of quiescent and active galaxies computed at $\log \mstare/\msune=11$ as a function of the virial factor \fvir. \emph{Right}: Displacement in $\log \mbhe$ between the $\log \mbhe-\log \sise$ relation of quiescent and active galaxies computed at $\log \sise/\kmse=2.2$ as a function of the virial factor \fvir. Solid red and long-dashed blue lines refer to the sample of \citet{HoKim14} and \citet[][only reverberation-based AGN]{Remco16}, as labelled. The top and lower panels adopt the observed (grey band) and unbiased (yellow bands) black hole relations shown in \figus2 and~3, respectively. The required virial factors are in general large for the observed relations ($\fvire>6$). For the $\log \mbhe-\log \mstare$ relation the \fvir\ factor is systematically higher by at least a factor of two, as expected in the presence of a bias in the sample of quiescent galaxies. The same virial factors of $\fvire\sim 3-5$ are instead consistent with the de-biased relations within $\lesssim 0.1$ dex uncertainty (see text for details). The vertical magenta dotted lines mark the reference virial factor of $\fvire=3.5$.
    \label{fig|FigureFvirSummary}}}
\end{figure*}

Last but not least, in line with local quiescent black holes, all the AGN data sets considered in this work lack low black hole masses in massive hosts.
The existence of a finite (though possibly large) scatter in the local scaling relations supports
the view of an underlying black hole-host galaxy correlation (a ``ridge'' as labelled by \citealt{Shankar16BH}), and disfavours the
idea that local black hole-host scaling relations mark the upper envelope of
a uniform distribution of black hole masses \citep[e.g.,][]{Bat10}.

\subsection{The \mbh-\sis\ relation of local active galaxies}
\label{subsec|localMbhSigma}

In \figu\ref{fig|FigureMbhSigmaAGN} we present a similar comparison between quiescent and active galaxies as in \figu\ref{fig|FigureMbhMstarAGN} but for the \mbh-\sis\ relation.
The open black squares and brown triangles are the early and late galaxies by \citet{SavorgnanGraham16} with velocity dispersions originally taken from Hyperleda database \citep{Paturel03} with a nominal aperture of $0.595$ kpc. We thus correct via \eq\ref{eq|veldispSersic} all their velocity dispersions to an aperture equal to one effective radius, with effective radii also self-consistently measured by \citet{SavorgnanGraham16}. A direct linear fit (see Appendix A in \citealt{Shankar17BH}) to the quiescent early-type sample yields (black long-dashed line)
\begin{equation}
\log \frac{\mbhe}{\msune}=8.51 + 5.09\,\left(\log \frac{\sigma_e}{\kmse}-2.3\right)\, ,
\label{eq|MbhSigma}
\end{equation}
with a scatter around $\sim 0.4$ dex (grey region).
\eq\ref{eq|MbhSigma} is close in both slope and normalization to \eq5 in \citet{KormendyHo}.
The fit to the early-type black hole population is very similar (black long-dashed line) as both early- and late-type galaxies are
less scattered in the \mbh-\sis\ relation \citep[e.g.,][and references therein]{Remco16}.
The Monte Carlo simulations by \citet{Shankar16BH} have revealed that assuming an underlying basic selection bias in the gravitational sphere of the central black hole yields an observed \mbh-\sis\ offset from the intrinsic/unbiased one by $\sim 0.5$ dex but with a similar slope, if the intrinsic scatter is $\lesssim 0.3$ dex (their Figure 9).
For our reference intrinsic/unbiased \mbh-\sis\ relation we thus choose to adopt \eq\ref{eq|MbhSigma} simply lowered in normalization by 0.5 dex (solid red line and yellow region).
For completeness, as in previous Figures, we also report in \figu\ref{fig|FigureMbhSigmaAGN} the elliptical sample of \citet[][open red triangles]{KormendyHo} which are fully consistent with the observed \mbh-\sis\ relation from the \citet{SavorgnanGraham16} sample of quiescent galaxies (black lines).

Following the format of \figu\ref{fig|FigureMbhMstarAGN}, retaining the AGN samples for which we have published velocity dispersions,
in the upper panels of \figu\ref{fig|FigureMbhSigmaAGN} we include the classical and pseudobulges reverberation-based black holes from
\citet[][filled red circles and green squares, respectively]{HoKim14} and the virial-based low-mass Seyferts from \citet[][orchid upside down triangles]{Mezcua18}, in the lower panels the megamasers from \citet[][filled cyan stars]{greene2016megamaser}, and the virial-based black holes and masers from
\citet[][blue diamonds and cyan circles for reverberation-based and masers AGN, respectively]{Remco16}. In the left panels we assume a uniform virial factor $\fvire=3.5$ and $\fvire=8$ in the right panels. It is evident that, at least for galaxies with velocity dispersions above $\log\sise/\kmse \gtrsim 1.9$, the former provides a better match to the unbiased relation, while the latter to the observed relation. More quantitative estimates of the appropriate virial scaling factors are given in \sect\ref{subsec|fvir}. Above $\log\sise/\kmse \gtrsim 1.9$, active black holes tend to follow a \mbh-\sis\ relation similar in slope to the quiescent samples, in line with what also claimed by \citet{Woo13AGN}. At lower velocity dispersions, below $\log\sise/\kmse \lesssim 1.9$,
the \citet[][orchid upside down triangles in the upper panels]{Mezcua18} sample tends to flatten out,
as already noted by \citet{Mezcua18}, and also tend to become more scattered. \citet{Baldassare16} on the other hand found that the lowest mass galaxies with AGN signatures in their sample fall within the extrapolations of the \mbh-\sis\ of quiescent black holes.
The main conclusions of this work anyway hold for galaxies with $\log\sise/\kmse \gtrsim 1.9$, i.e. with stellar masses $\log \mstare/\msune \gtrsim 10$ according to the mean \sis-\mstar\ relation of SDSS galaxies (long-dashed line in \figu\ref{fig|FigureSigmaMstar}). The data on both active and quiescent galaxies become much more sparse below this threshold.

\subsection{A distinct virial factor \fvir\ for the observed \mbh-\mstar\ and \mbh-\sis\ relations}
\label{subsec|fvir}

In the previous sections we found that assuming an underlying virial factor of $\fvire=3.5$ provides a reasonable match to both the unbiased \mbh-\mstar\ and \mbh-\sis\ relations put forward by \citet{Shankar16BH}. On the other hand, larger values of \fvir\ are required to broadly match the observed scaling relations of black holes in quiescent galaxies. In particular, there is evidence for the need of a particularly large \fvir\ when comparing with the observed \mbh-\mstar\ relation (right panels of \figu\ref{fig|FigureMbhMstarAGN}). In \figu\ref{fig|FigureFvirSummary} we provide a more thorough investigation into the distributions of virial factors \fvir\ required to match both the observed (upper panels) and unbiased (lower panels) relations. For this purpose we choose the two reverberation-based data sets of classical/pseudobulges by \citet[][red solid lines]{HoKim14} and the virial-based AGN by
\citet[][blue long-dashed lines]{Remco16}. Both samples share a broad coverage in stellar mass and velocity dispersion, and are not biased with respect to the \sis-\mstar\ relation of SDSS galaxies (\figu\ref{fig|FigureSigmaMstar}). For each AGN sample and a given input $\fvire$ we then perform a linear fit to the \mbh-\sis\ and \mbh-\mstar\ relations and compute their offsets with respect to the analogous relations for quiescent galaxies. Our methodology, which is slightly different from the one put forward by, e.g., \citet[][their \eq3]{Onken04}, allows to determine the offsets between the active and quiescent populations in a given bin of stellar mass or velocity dispersion, particularly relevant when the slopes tend to be different, as in the \mbh-\mstar\ plane (\figu\ref{fig|FigureMbhMstarAGN}).

The left panels of \figu\ref{fig|FigureFvirSummary} show the
displacement in $\log \mbhe$ between the $\log \mbhe-\log \mstare$ relation of quiescent and active galaxies computed at $\log \mstare/\msune=11$ as a function of the virial factor \fvir.
Both samples suggest that, in agreement with \figu\ref{fig|FigureMbhMstarAGN}, only virial factors $\fvire\gtrsim 15$ tend to align the AGN with the quiescent galaxies within $\Delta \log \mbhe \lesssim 0.1$ dex in the \mbh-\mstar\ relation. The reference virial factor of $\fvire=3.5$ (vertical magenta dotted lines in \figu\ref{fig|FigureFvirSummary}), in particular, would generate an offset of $\Delta \log \mbhe \sim 0.8$ dex, in order for the reverberation/virial-based AGN to line up to the relation observed in quiescent ellipticals. Note that we deliberately chose to compute the displacements at relatively high stellar masses $\log \mstare/\msune=11$ as at lower masses the match would require even higher \fvir\ to bring AGN in line with the relation of quiescent galaxies, being the former steeper than the latter (see \sect\ref{subsec|localMbhMstar}).
In the right panels of \figu\ref{fig|FigureFvirSummary} we instead plot as a function of the virial factor \fvir\ the displacement in $\log \mbhe$ between the $\log \mbhe-\log \sise$ relation of quiescent and active galaxies computed at $\log \sise/\kmse=2.2$, which is the mean velocity dispersion roughly corresponding to $\log \mstare/\msune=11$, following the SDSS \sis-\mstar\ relation (\figu\ref{fig|FigureSigmaMstar}). In this case the match between AGN and quiescent galaxies within $\Delta \log \mbhe\lesssim 0.1$ dex in the \mbh-\sis\ relation is reached for $\fvire\gtrsim 7$, roughly a systematic factor of two lower than the $\fvire$ required to match the \mbh-\mstar\ relation.

This systematic difference in virial factors can be easily interpreted within, and in fact taken in support of, the framework of the selection bias considered by \citet{Shankar16BH}.
As discussed with respect to \figu\ref{fig|FigureSigmaMstar}, this bias induces the selection of sources with velocity dispersions on average higher
than regular counterpart SDSS galaxies of similar stellar mass. In turn, higher velocity dispersions
would imply selecting on average higher mass black holes, given the strong dependence $\mbhe\propto \sise^{4-5}$.
Ultimately, this selection bias is expected to artificially increase the normalization of the black hole scaling relations, in particular of the \mbh-\mstar\ relation which is a direct reflection of the \sis-\mstar\ relation. In practise, the presence of such a bias requires a mean \fvir\ factor systematically higher, by at least a factor of two, in the \mbh-\mstar\ than in the \mbh-\sis\ relation. On the other hand, the \emph{same} virial factors of $\fvire \sim 3-4$ are consistent with both the intrinsic relations within $\lesssim 0.1$ dex uncertainty (bottom panels of \figu\ref{fig|FigureFvirSummary}). A $\fvire \sim 3-4$ also do not show the discrepancy between the masers and the reverberation-based black holes which appears instead when a higher \fvir\ is adopted (compare the cyan stars and cyan circles in the middle, right panel of \figu\ref{fig|FigureMbhMstarAGN}).

\subsection{Residuals in the active black hole scaling relations} %in the \mbh-\mstar\ and \mbh-\sis\ relations
\label{subsec|residuals}

\begin{figure*}
    \center{\includegraphics[width=\textwidth]{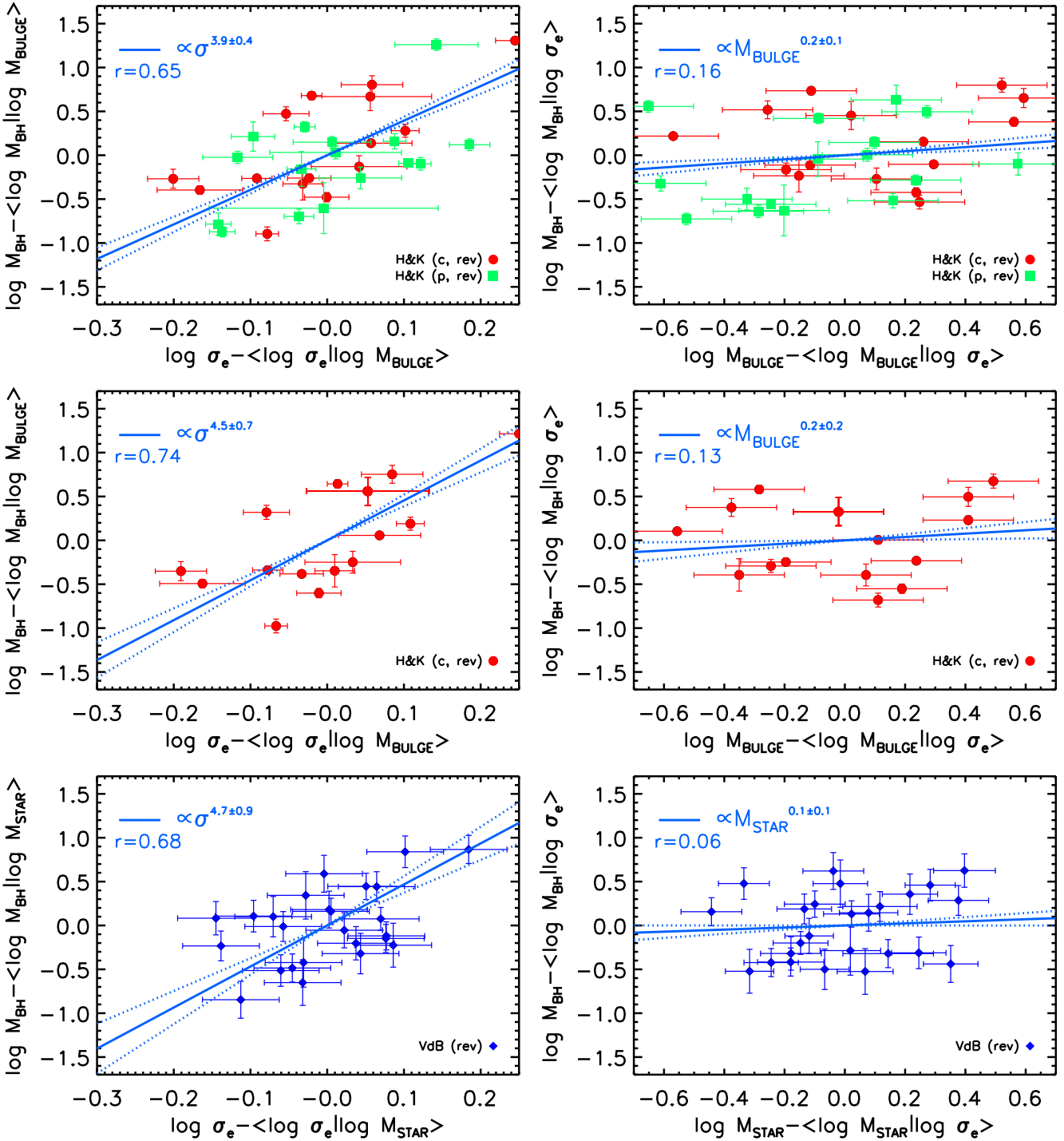}
    \caption{Correlations between residuals from the observed scaling relations, as indicated. Red circles and green squares in the top panels show, respectively, classical and pseudobulges from \citet{HoKim14}. The middle panels show the subsample of classical bulges. The blue filled diamonds in the bottom panels are the reverberation-based data from \citet{Remco16}.
    Correlations with velocity dispersion appear to be much
    stronger (left panels; Pearson coefficient $r\sim 0.6-0.7$) than those with bulge/total stellar mass (right panels; $r\lesssim 0.1-0.2$),
    in line with that observed for local samples of quiescent supermassive black holes with dynamical mass measurements.
    \label{fig|FigureResiduals}}}
\end{figure*}

It has been discussed by a number of groups \citep{Bernardi07,Shankar16BH,Remco16,Barausse17,Shankar17BH} that
velocity dispersion is a galactic property linked to black hole mass more fundamentally in terms of residuals than any other one, including stellar/bulge mass, light profile concentration, or effective radius. Residuals in pairwise correlations \citep[][]{ShethBernardi12} are in fact an effective methodology to probe underlying relations among variables.
The left and right panels of \figu\ref{fig|FigureResiduals} report the $\Delta(\mbhe|\mbulgee)$ vs $\Delta(\sise|\mbulgee)$ and
$\Delta(\mbhe|\sise)$ vs $\Delta(\mbulgee|\sise)$, where
\begin{equation}
\Delta(Y|X)\equiv\log Y-\langle \log Y|\log X \rangle \,
\label{eq|resid}
\end{equation}
is the residual in the $Y$ variable (at fixed $X$) from the log-log-linear fit of $Y(X)$ vs $X$, i.e. $\langle \log Y|\log X \rangle$.

For this test, analogously to what performed in \figu\ref{fig|FigureFvirSummary},
we make use of the reverberation-based \citet{HoKim14} and virial-based \citet{Remco16} samples which
share a broad coverage in stellar mass and velocity dispersion and are not biased with respect to the \sis-\mstar\ relation of SDSS galaxies.
For the \citet{HoKim14} sample in particular, we show results using \emph{bulge} rather than total stellar masses, being the former more physically related to central velocity dispersion.
We assume a constant $\fvire=3.5$ for all sources in both samples.
To produce residuals we follow the procedure outlined in \citet{Shankar16BH} and \citet{Shankar17BH}. For each residual we run 200 iterations, and at each iteration we eliminate two random objects from the original sample. From the full ensemble of realizations, we measure the mean slope and its 1$\sigma$ uncertainty.

\figu\ref{fig|FigureResiduals} clearly shows that black hole mass is strongly correlated with velocity dispersion at fixed bulge stellar mass with a Pearson coefficient $r=0.65$ (top left panel), while the correlation with bulge mass is negligible with $r=0.16$ at fixed velocity dispersion (right top panel). When computing residuals for only classical bulges (red circles, middle panels), the correlation with velocity dispersion becomes even more marked with a Pearson coefficient of $r=0.74$ while the one with bulge mass further decreases to $r=0.13$.
Adopting total stellar masses in the \citet{HoKim14} sample and/or their original values for \fvir\ would produce even stronger residuals with velocity dispersion, further corroborating our results.
The lower panels of \figu\ref{fig|FigureResiduals} show that the residuals extracted from the reverberation-based sample from \citet{Remco16} still point to strong correlation with velocity dispersion ($r=0.68$) and a negligible one with (total) stellar mass ($r=0.06$).

These findings fully confirm and extend to local AGN the dominance of velocity dispersion in pairwise scaling relations put forward by \citet{Bernardi07} and \citet{Shankar16BH} for local quiescent black holes. The total slope of the \mbh-\sis\ relation in AGN can be estimated
as $\mbhe\propto\sigma^{\beta}\mstare^{\alpha}\propto\sigma^{\beta+\alpha\,\gamma}$, where $\gamma$ comes from $\mstare\propto\sigma^{\gamma}$. Since SDSS
galaxies tend towards $\gamma\approx 2.2$ \citep{Shankar17BH}, and the residuals in \figu\ref{fig|FigureResiduals} yield $\beta\sim 4-5$ and $\alpha\sim 0.1-0.2$, one obtains a total dependence of $\mbhe\propto\sigma^{4.2-5.5}$, consistent with models of black hole growth being regulated by energy/momentum-driven AGN feedback \citep[e.g.,][]{SilkRees,Fabian99,King03,Granato04}.

\section{Discussion and Conclusions}
\label{sec|discu}

%The first result of the present work is that the samples of local ($z<0.3$) AGN considered here with measured velocity dispersions and stellar masses are consistent, on  average, with those from the SDSS galaxies (\figu\ref{fig|FigureSigmaMstar}), and fall well below the local scaling relations of quiescent galaxies, especially with respect to the \mbh-\mstar\ relation (\figu\ref{fig|FigureMbhMstarAGN}).

The first result of the present work is that the samples of local ($z<0.3$) AGN considered here with measured velocity dispersions and stellar masses are consistent, on average, with those from the SDSS galaxies (\figu\ref{fig|FigureSigmaMstar}).  This is in stark contrast to samples of quiescent black holes, which are biased with respect to the full SDSS sample \citep[][]{Bernardi07,Shankar16BH}. The second result concerns the factor \fvir\ that is adopted to normalise black hole masses in reverberation-based AGN samples \citep[e.g.,][]{Peterson93}. Values of order $\fvire\sim 15$ are required to match the selection-biased \mbh-\mstar\ relation of quiescent galaxies, whereas $\fvire\sim 3.5$, as derived from direct modelling of the broad line region \citep[e.g.,][and references therein]{Williams18}, is required to match the \citet{Shankar16BH}'s estimate of the intrinsic relation for quiescent galaxies (\figu\ref{fig|FigureMbhMstarAGN}). Water masers, with black hole masses independent of any virial factor, also fall within the scatter of the unbiased relations by
\citet[][cyan symbols in \figus\ref{fig|FigureMbhMstarAGN} and \ref{fig|FigureMbhSigmaAGN}]{Shankar16BH}.

A number of other groups had also found significant discrepancies between AGN samples, with virial factors \fvir\ calibrated against some renditions of the \mbh-\sis\ relation, and the observed \mbh-\mstar\ relation of quiescent black holes \citep[e.g.,][]{busch2014low,Dasyra07,Kim08,Sarria10,Falomo14,ReinesVolonteri15,greene2016megamaser,Bentz18}, though without offering a convincing explanation for the nature of this puzzling offset. \citet{ReinesVolonteri15}, in particular, carried out a detailed work to compare one of the largest local samples of broad-line AGN (244 sources) with virial black hole mass estimates, with the samples of quiescent galaxies from \citet{KormendyHo}, as (re)proposed in the bottom panels of our \figu\ref{fig|FigureMbhMstarAGN}. After applying appropriate corrections to different total stellar mass estimates, they concluded that the substantial discrepancy ($\sim 1.2$ dex) in the normalization between the active and quiescent samples cannot be ascribed to only measurement errors. Indeed, they found that reconciling the two samples would require average virial factors of the order of $\fvire \gtrsim 40$, which is significantly beyond the typical uncertainties in \fvir. \citet[][]{Bennert11Local} also found evidence for a mean discrepancy between local active and inactive galaxies in terms of spheroidal luminosity (left upper panel of their \figu17), though it tends to disappear when converting to stellar masses (middle upper panel of their \figu17) most probably due to their specific choices of mass-to-light ratios.

We are thus not the first to report on an apparent discrepancy between scaling relations of active and quiescent local black hole samples \citep[e.g.,][]{ReinesVolonteri15}. However, our work offers new insights into the origin of this discrepancy.
In particular, it highlights the important role played by angular resolution-related selection effects on the quiescent sample.
We showed in fact that the mean \fvir\ required to match the selection-based \mbh-\mstar\ relation of quiescent black holes is systematically higher by a factor $\sim 2-3$ than that needed to match the \mbh-\sis\ relation. However, once the selection effect has been accounted for, values of $\fvire\sim3.5$ yield agreement with both relations. \citet[][]{HoKim14} also found evidence for a systematic difference in virial factors. They claimed that classical, more massive bulges, required $\fvire\sim 6.3$ to match the $\mbhe-\sise$ relation of \citet{KormendyHo}, and a virial factor of $\fvire\sim 9$ ($\sim 0.4$ dex higher) to match the \mbh-\mbulge\ relation.

Previously, before the black hole mass revisions outlined in \citet[][and references therein]{KormendyHo}, lower values of the \fvir\ factors had been reported in the literature (see \sect\ref{sec|intro}). For example, when comparing to the \citet{Woo13} \mbh-\sis\ relation of quiescent black holes, \citet{Grier13} retrieved a mean $\fvire \sim 4.3$. \citet{Shankar16BH} reported that mean virial factors calibrated on the observed \mbh-\sis\ relation should be reduced by a factor of $\sim 3$ to account for selection bias (see also \figu\ref{fig|FigureMbhSigmaAGN}). They thus suggested that the mean virial factor could be of order unity based on the \citet{Grier13} estimate of $\fvire \sim 4$. However, we here showed, in broad accordance with \citet{HoKim14}, that a more proper comparison with the most recent scaling relations of quiescent black holes suggests $\fvire \sim 10-12$ for a close match to the \mbh-\sis\ relation (upper right panel of \figu\ref{fig|FigureMbhSigmaAGN}), which would then imply a scaled $\fvire \sim 3.5-4$ to match the unbiased relation (left panels of \figu\ref{fig|FigureMbhSigmaAGN}), in nice agreement with the independent estimate of $\fvire \sim 3.5$ by, e.g., \citet[][]{Grier17} and \citet[][]{Williams18}. Finally, the correlation between \mbh-\mstar\ of the masers is more consistent with that of the reverberation-based black holes if a $\fvire \sim 3-4$ is adopted instead of a higher value.

It is very unlikely, also given the great diversity of AGN samples collected here, that the observed large offsets in black hole mass with respect to the observed relations of local quiescent black holes is a mere consequence of ``infant'' black holes still growing towards their final mass \citep[e.g.,][]{Mathur12}, especially in the local Universe, vastly dominated by AGN with low characteristic Eddington ratios \citep[e.g.,][]{Kauffmann09,Shankar13acc}.
These findings instead lend further support to the presence of a strong bias in the local quiescent black hole sample, possibly caused by the limited telescope resolution power that artificially increases the normalization of the \sis-\mstar\ relation along with other scalings \citep[e.g.,][]{Bernardi07,Shankar16BH,Shankar17BH}. This bias has important consequences on a number of fronts, from implications for seeds and intermediate mass black holes \citep[e.g.,][]{GrahamReview15,Shankar16BH,Mezcua17review,Pacucci18}, gravitational waves \citep[e.g.,][]{Sesana16,Taylor17,Chen18,Joshi18}, to radiative/kinetic efficiencies (Shankar et al. 2019, submitted).

\citet{ReinesVolonteri15} pointed out that the quiescent subsample of later-type galaxies, especially pseudobulges, with dynamically-measured black holes tend to overlap with the black hole scaling relations in active galaxies. We do find a similar trend with the (few) late-type galaxies in the \citet{SavorgnanGraham16} sample overlapping with our intrinsic black hole scaling relations (\figus\ref{fig|FigureMbhMstarAGN} and \ref{fig|FigureMbhSigmaAGN}). In our interpretation, the displacement, or lack thereof, between the relations of active and quiescent black holes can be explained by how much the host galaxies are outliers in the SDSS \sis-\mstar\ relation (\figu\ref{fig|FigureSigmaMstar}). Later-type galaxies, usually characterized by lower velocity dispersions at fixed stellar mass, will always be less biased than earlier-type galaxies, thus naturally appearing more consistent with the intrinsic scaling relations of black holes.

We stress that all the results presented here, including \eqs\ref{eq|IntrinsicMbhMstar} and \ref{eq|MbhSigma}, are mostly applicable to galaxies with (total) stellar mass $\log \mstare/\msune \gtrsim 10$ and velocity dispersion $\log \sise/\kmse\ \gtrsim 1.9$. Although our AGN samples and SDSS data become rapidly more sparse and/or less robust below this limit, some efforts to probe black hole scaling relations at lower stellar masses have been attempted in recent years. \citet[see, e.g.,][and references therein]{Graham18Late} suggest that a steepening should be present in the \mbh-\mstar\ relation at low stellar masses, while a flattening in the \mbh-\sis\ relation seems to be more consistent with the data according to \citet{Mezcua18} and \figu\ref{fig|FigureMbhSigmaAGN} (but see also \citealt[][]{Baldassare16}). In the context of further probing the correlations of active black holes at lower stellar masses, we also considered six low-mass galaxies from the Sloan Digital Sky Survey-IV Mapping Nearby Galaxies with AGN signatures \citep[][]{Penny18}. We computed stellar masses for these galaxies from their $r-i$ colours and the mass-to-light ratios from \citet{Bell03SEDs}, decreased by 0.1 dex to convert to a Chabrier IMF. Velocity dispersion were corrected to an aperture of one effective radius using \eq\ref{eq|veldispSersic}, in which the S\'{e}rsic index was set to $n_{\rm Ser}=2$ (this choice does not have any impact on our results). Black hole masses were then inferred from \eq\ref{eq|MbhSigma} lowered by 0.5 dex to convert to the intrinsic \mbh-\sis. We found the median black hole mass competing to a median host stellar mass of $\log \mstare/\msune \sim 9.7$, to be $\log \mbhe/\msune \sim 5.3$, which should be regarded more as an upper limit, as three galaxies have only assigned upper limits to their velocity dispersions. The median black hole mass in these dwarfs lies nicely within the $1 \sigma$ of the intrinsic \mbh-\mstar\ relation reported in \figu\ref{fig|FigureMbhSigmaAGN}. We have also compared with the sample of dwarfs included in Table 3 of \citet{ReinesVolonteri15}, which yield, once converted to our reference mass-to-light ratio via \eq\ref{eq|ZibettiBell}, a mean black hole mass of $\log \mbhe/\msune \sim 5.6\pm0.5$ at $\log \mstare/\msune \sim 9.7$. The latter sample of dwarfs is $\sim 2 \sigma$ above the extrapolation of \eq\ref{eq|IntrinsicMbhMstar}, possibly suggesting that a drop in black hole masses as marked as predicted by the \emph{extrapolation} of
our \eq\ref{eq|IntrinsicMbhMstar} may be inconsistent with present data. In any event probing the \mbh-\mstar\ correlation at such low stellar masses falls beyond the parameter space probed by the Monte Carlo simulations performed by \citet{Shankar16BH} which, as already stressed above, were based on relatively massive galaxies with $\log \mstare/\msune \gtrsim 10$. What is more relevant to emphasize in the context of the present paper is that all the dwarfs considered here fall an order of magnitude of more below the extrapolation of the \emph{observed} \mbh-\mstar\ relation of quiescent black holes (grey region in \figu\ref{fig|FigureMbhMstarAGN}).

It is relevant to note that most of the data adopted here are extracted from samples of Type 1 AGN. We note, however, that recent work has proven Type 2 AGN to be characterized, if anything, by even lower scaling relations \citep[e.g.,][]{Ricci17SR} which would further strengthen our claim for a bias in the \mbh-\mstar\ relation of dynamically measured quiescent supermassive black holes.

We have finally investigated into AGN samples at higher redshifts, with stellar masses/black hole masses derived from spectral energy decomposition analysis and single epoch spectroscopy \citep[e.g.,][]{Chang17AGN,delvecchio2017vla,Rowan17}. In all cases we find the black holes to lie substantially below the local observed \mbh-\mstar\ relation. However, probing the presence of biases in the \mbh-\mstar\ relation at higher redshifts becomes increasingly less robust due to the possibility of sampling growing black holes, and/or due to intrinsic evolutionary effects in the \mbh-\mstar\ relation itself, and/or due to severe flux limitations \citep[e.g.,][]{Bongiorno14}.

\vspace{0.15cm}
\noindent To conclude, in this work we showed that:
\begin{enumerate}
  \item AGN hosts follow the \sis-\mstar\ correlation of SDSS galaxies, in stark contrast to quiescent galaxies with dynamical mass measurements of their black holes (\figu\ref{fig|FigureSigmaMstar}). This behaviour can be in large part explained by the selection effect imposed on the sample of quiescent galaxies by the limited telescope resolution power, which forces to preferentially target more massive, larger velocity dispersions black holes to allow for resolved dynamical mass modelling.
  \item When adopting mean virial factors of $\fvire=3.5$ from direct modelling of the broad line region, distinct data sets of local active galaxies with accurate galaxy mass measurements all point to a \mbh-\mstar\ relation steeper and with a much lower normalization than the selection biased one observed for local quiescent black holes with dynamical mass measurements (\figu\ref{fig|FigureMbhMstarAGN}).
  \item The same data sets of local active galaxies with also available velocity dispersion measurements, all point to a \mbh-\sis\ relation, at least at $\log \sise/\kmse \gtrsim 1.9$, with a similar slope but offset by a factor of $\sim 3$ from the selection biased one observed for local quiescent black holes with dynamical mass measurements (\figu\ref{fig|FigureMbhSigmaAGN}).
  \item The mean virial factors required for the AGN samples with reverberation-based black hole measurements to match the high-mass end of, respectively, the \mbh-\sis\ and \mbh-\mstar\ relations of quiescent galaxies within $\lesssim 0.1$ dex accuracy are $\fvire\gtrsim 7$ and $\fvire\gtrsim 15$, substantially higher than the canonical $\fvire\sim 3.5$  (\figu\ref{fig|FigureFvirSummary}). This systematic offset in \fvir\ can be naturally explained within the framework of the same selection effects in the local sample of quiescent black holes, which tend to be more effective in the \mbh-\mstar\ rather than the \mbh-\sis\ plane.
  \item The residuals in the scaling relations of AGN show clear evidence for a strong intrinsic dependence of black hole mass on velocity dispersion but not on (bulge or total) stellar mass (\figu\ref{fig|FigureResiduals}), in line with what observed in the sample of local inactive black holes \citep{Shankar16BH}. These results support the finding of \citet{Bernardi05} and \citet{Bernardi11b} who showed that \sis\ is the main driver of scaling relations with \mstar.

\end{enumerate}

\section*{Acknowledgments}

We warmly acknowledge Daniel Asmus, Misty Bentz, Angela Bongiorno, James Collinson, Ivan Delvecchio, Sebastian H\"{o}nig, Michael Koss, Mirko Krumpe, Isabella Lamberti, Achille Nucita, Michael Rowan-Robinson, and Andreas Schulze for a number of useful discussions and for kindly sharing their data. FS acknowledges partial support from a Leverhulme Trust Research Fellowship. VA acknowledges funding from the European Union's Horizon 2020 research and innovation programme under grant agreement No 749348. MM acknowledges support from the Spanish Juan de la Cierva program (IJCI-2015-23944). FR acknowledges support from FONDECYT Postdoctorado 3180506 and CONICYT project Basal AFB-170002. FP acknowledges support from the NASA Chandra award No. AR8-19021A. We thank the referee for useful suggestions that helped to improve the clarity of the paper.
\bibliographystyle{mn2e_Daly}
\bibliography{MNRASfvirN2.bbl}
%\bibliography{../RefMajor_Rossella}

\appendix
\section{AGN samples used in the main text}
\label{Appendix}

We here describe how we have corrected the stellar and black hole masses to a {\it homogeneous} system in our reference AGN samples.
As in the main text, we here always refer to \fvir\ as the virial factor calibrated on the line dispersion (i.e., second moment of the line profile)
of H$\beta$ \citep[e.g.,][]{Peterson04}.
For virial black hole masses calibrated adopting the full width at half maximum (FWHM) of the H$\alpha$ line,
we will instead denote the virial factors with the symbol $\epsilon$, following the convention put forward by \citet{Onken04}, which is roughly a factor of a few lower than \fvir.
All the AGN samples considered in this work adopt as a reference the line dispersion of the broad
H$\beta$ emission line, except for the samples by \citealt{ReinesVolonteri15} and \citealt{Mezcua18}, who adopt the FWHMs of the H$\alpha$ lines.
\citet{Grier17} have fitted the continuum H$\beta$ emission-line spectra finding that
$\fvire\sim 3.5\, (\log \fvire=0.54\pm 0.17)$ and $\epsie\sim 1.5\, (\log \epsie=0.18\pm 0.23)$ when adopting the line dispersion and FWHM of the H$\beta$ line, respectively.
Following \citet{Grier17}, in this work we use as our (mean) reference virial factors $\fvire=3.5$ and $\epsilon=1.51$.
When adopting larger values of \fvir\ (e.g., as in the right panels of \figu\ref{fig|FigureMbhMstarAGN}),
we increase our reference $\epsilon$ proportionally to the ratio $\fvire/3.5$.
We here assume the same equivalent virial factors apply to both the H$\alpha$ and the H$\beta$ emission lines.
This is corroborated by \citet{Greene05}, who found a tight linear correlation between the H$\alpha$ and H$\beta$ FWHMs
with only a relatively small offset of $\lesssim 0.05$ dex between the two lines (their \figu3), and also more recently by \citet[][their \figu1]{Ricci17b}.
This offset would
correspond to a systematic difference in the implied black hole masses of $\lesssim 25\%$,
which is much lower than the level of systematics discussed in this work. We now provide a brief description of each of the AGN samples considered in this work:\\
\noindent (1) \citet{ReinesVolonteri15} carried out a study of the black hole mass-host galaxy stellar mass relation of 262 broad-line AGN at $z \leq 0.055$, from the SDSS. Stellar masses were derived from the \citet{Zibetti09} colour-dependent mass-to-light ratios. After correcting for a different IMF, \citet{ReinesVolonteri15} showed that the \citet{Zibetti09} stellar masses present a mass-dependent offest with respect to
the \citet{Bell03SEDs} stellar masses (their \figu6, upper panel) which we approximate as
\begin{multline}
\log M_{\rm star, Bell}=1.793+0.845\times\log M_{\rm star, Zib}\, ,
    \label{eq|ZibettiBell}
\end{multline}
which provides, in the range of interest here $9<\log \mstare<11.5$, a median offset of $\sim 0.21$ dex, as measured by \citet{ReinesVolonteri15}.
\citet{ReinesVolonteri15} calculated black hole masses from single-epoch virial estimators based on the H$\alpha$
emission line and luminosity, with a mean virial coefficient $\epsilon=1.075$ \citep{Grier13,Onken04}, which we correct to our reference value of $\epsilon=1.51$.\\
\noindent (2) \citet{HoKim14} collected 43 reverberation mapping AGN with S\'{e}rsic \citep{Sersic63} bulge-to-total decompositions on \emph{Hubble Space Telescope} imaging. Bulge masses were then computed from the $B-R$ colour-dependent \ML\ from \citet{Bell03SEDs}, and converted to total stellar masses using their tabulated values of bulge-to-total fractions.
\citet{Bell03SEDs} adopted a ``diet Salpeter'' IMF, which we corrected to our
reference Chabrier IMF subtracting 0.1 dex to all stellar masses \citep[see Table 2 in][]{Bernardi10}. We assign to all their stellar masses a typical statistical error of 0.15 dex \citep[][]{Bernardi12}.
\citet{HoKim14} took bulge velocity dispersions mostly from \citet{Nelson04} and \citet{Woo13AGN},
who calibrated their velocity dispersions specifically at the effective radius. Other measurements adopted by \citet{HoKim14}
are from, e.g., \citet{NelsonWhittle}, \citet{Ferrarese01}, or \citet{Woo10AGN}, who extracted spectra from
2'' slits or a few arcsecond square apertures centred on the nucleus, similar to the aperture adopted in SDSS galaxies.
%1.5/2.5''$\times$2/3'' and 2''$\times$4''
As discussed by \citet{Ferrarese01} and \citet{Nelson04},
these velocity dispersion measurements are effectively carried out at several kiloparsecs from the centre, probe the stellar bulge potential,
and show relatively flat profiles \citep[e.g.,][]{MerrittFerrarese01,KormendyHo}.
We thus treat all their measured velocity dispersions as equivalent to those in our SDSS sample within one effective radius.
As a further test, we also checked that fully consistent results are found in both the \sis-\mstar\ relation
(\figu\ref{fig|FigureSigmaMstar}) and residuals (\figu\ref{fig|FigureResiduals}),
when adopting, wherever possible, velocity dispersions from the Hyperleda database \citep{Paturel03}.
\citet{HoKim14} divided their sample into classical and pseudobulges based on their S\'{e}rsic index, for which they calibrated, respectively,
$\fvire=6.3$ and $\fvire=3.2$ (based on the H$\beta$ line dispersion) when compared to the local \mbh-\sis\ relation of inactive galaxies from \citet{KormendyHo}.
We instead adopt for all black hole masses in their sample a constant $\fvire=3.5$.\\
\noindent (3) \citet{busch2014low} performed aperture photometry and bulge-to-disc decompositions on near-infrared $J-$, $H-$, and $K-$band images
of 20 low-luminosity type-1 quasars at $z\leq0.06$ from the Hamburg/ESO survey. Stellar masses were
calculated from parametric models based on inactive galaxy colours and using the \citet{Bell03SEDs} mass-to-light ratios, which we
again correct to our reference Chabrier IMF subtracting 0.1 dex.
Black hole masses were derived from the H$\beta$ line dispersion virial-based sample of \citet{Schulze09} with $\fvire=3.85$ \citep{Collin06},
which we rescale to $\fvire=3.5$.\\
\noindent (4) \citet{greene2016megamaser} presented black hole masses and stellar velocity dispersion measurements for several local megamaser early-to-mid-type spiral galaxies, including those from the \citet{Saglia16} sample. Black hole masses were derived by fitting a Keplerian rotation curve to the positions and velocities of the maser spots. Using SDSS data, stellar masses were obtained from the \citet{Bell03SEDs} mass-to-light ratios, for which we subtract a constant 0.1 dex.
Velocity dispersions were extracted from spectra within a 2'' aperture and were measured from the first two moments of the line-of-sight velocity dispersion.
Analogously to the \citet{Ferrarese01} measurements, even for this sample we do not apply any correction to the velocity dispersions when comparing to the SDSS $\sigma_e$.\\
\noindent (5) \citet{Mezcua18} studied 127 low-mass Seyfert 1 galaxies with measured stellar masses from the SDSS spectral measurements and
\citet{BruzualCharlot03} libraries with a Chabrier IMF, which are fully consistent with the stellar masses from \citet{Kauffmann04} and systematically lower than those calibrated using \citet{Bell03SEDs} by $\sim 0.25$ dex \citep[e.g.,][]{Bernardi16ML}.
Black hole masses, derived from the H$\alpha$ virial relations of \citet[][and references therein]{Woo15} with a mean virial factor of $\epsie=1.12$ and $\epsie=0.75$, respectively, are converted to our reference $\epsie=1.51$.
Velocity dispersions were extracted from SDSS and corrected for an aperture within one (exponential) effective radius via the usual scaling
\begin{equation}
\sigma(R_e)=\sigma_{\rm SDSS}(1.5")\times\left(\frac{1.5"}{R_e}\right)^{\gamma}\, ,
    \label{eq|sigmaGamma}
\end{equation}
but using a S\'{e}rsic index-dependent slope \citep{Bernardi17IMF}
\begin{multline}
\gamma=-0.016366723+0.019136886\times n_{\rm Ser}\\
-0.0010648890\times \left(n_{\rm Ser}\right)^2\, .
    \label{eq|veldispSersic}
\end{multline}
We set the S\'{e}rsic index $n_{\rm Ser}=1$ in \eq\ref{eq|veldispSersic}, as the vast majority of the \citet{Mezcua18} sample is heavily disc-dominated. Assuming a larger value for the S\'{e}rsic index such as $n_{\rm Ser}=2$ has a negligible effect on our results.\\
\noindent (6) \citet{Remco16} compiled one of the largest samples of active and inactive supermassive black holes from the literature, counting 230 sources with black hole mass measurements from stellar dynamics, gas or CO dynamics, masers and reverberation mapping-based relations. In what follows, we will only consider the active galaxies from \citet{Remco16} from masers and reverberation-based estimates.
The latter are extracted from the original sample by \citet{Bentz15} who adopted a mean $\fvire=4.31$ from \citet[][H$\beta$ line dispersion]{Grier13} which we rescale to $\fvire=3.5$. \citet{Remco16} adopted, wherever possible, the closest approximation to $\sigma_e$ for the velocity dispersions (to which we assign typical statistical errors of 0.05 dex \citealt[][]{Bernardi10}), and uniformly applied S\'{e}rsic fits to the 2MASS $K_s$ band photometry. All galaxies in this sample, which excludes the most massive galaxies,
are well resolved and detected in 2MASS according to \citet{Remco16}.
Following \citet{Remco16}, galaxy luminosities were then converted to stellar masses via the velocity dispersion-dependent
mass-to-light ratio from \citet{KormendyHo}
\begin{equation}
\frac{M_{\rm star}}{L_K}=K_0\left(\frac{\sigma_e}{166\, \kmse}\right)^{0.45}\, \frac{M_{\rm \odot}}{L_{\rm \odot,K}}\, .
    \label{eq|MLfromKH}
\end{equation}
\citet{KormendyHo} specify that the constant in \eq\ref{eq|MLfromKH} should be $K_0\sim1$ for the total \emph{dynamical} mass-to-light ratios, and further detail that, following \citet{Into13},
it should be reduced to $K_0\sim 0.76$ when considering, as in this work, only the stellar component. We thus adopt the latter value for $K_0$.
We further subtract a constant 0.05 dex from all stellar masses to account for the heavier \citep{Kroupa01} IMF adopted by \citet{Into13}.
In the end, we find that galaxies with a Chabrier IMF and $\sigma_e=166\, \kmse$ have an average $M_{\rm star}/L_K\sim0.676$, which is in excellent
agreement with the mean $M_{\rm star}/L_K=0.67$ calculated by \citet{Longhetti09}. Indeed, we checked that equivalent results are found throughout
even ignoring the (weak) dependence on velocity dispersion in \eq\ref{eq|MLfromKH}. We note that the \citet{Remco16} maser subsample also includes the sources from \citet{greene2016megamaser} but with independent measurements of the host stellar masses so we will still consider them as fully independent estimates.\\
\noindent (7) \citet{Bentz18} recently collected a sample of 37 active galaxies with high-resolution optical images from the \emph{Hubble Space Telescope} with reverberation-based black hole masses with an average $\fvire=4.3$. Total luminosities are derived from S\'{e}rsic fits, though with a S\'{e}rsic index capped at $n_{\rm Ser}\le4$. We consider their mass-to-light ratios from \citet{BellJong01} who adopted a diet Salpeter. However, in this specific instance we do not substract their stellar masses by 0.1 dex to convert to a Chabrier IMF as \citet{Bell03SEDs} showed (their Figure 20, left panel) that at fixed optical galaxy colour the \citet{BellJong01} \ML\ are already systematically underestimated by (at least) $\sim 0.1$ dex with respect to those from \citet{Bell03SEDs}.

Other AGN samples are available in the local Universe but are less suited to our purposes. For example,
\citet{nucita2017catalog} presented X-ray AGN at $z\leq0.055$
with black hole masses derived from the X-ray/radio fundamental plane relation \citep{MerloniFP}.
However, the latter has already been shown to be biased, probably tracing the upper envelope of a much broader distribution \citep[e.g.,][]{LaFranca10,Bonchi13,Baldi18,Mezcua18}. \citet{koss2017bat} presented the first catalogue and data release of the Swift-BAT AGN Spectroscopic Survey. However, their stellar masses were obtained by combining the SDSS Petrosian photometry with the \citet{BlantonRoweis} mass-to-light ratios both of which contribute to yield stellar masses a factor of $\gtrsim 2$ lower than the updated measurements adopted here \citep[e.g.,][]{Bernardi10,Bernardi13}.  Nevertheless, we checked that after correcting stellar masses by a factor of two the \citet{koss2017bat} sample lies well below the observed \mbh-\mstar\ relation of quiescent galaxies.  Interestingly, \citet[][see also \figu8 in \citealt{Krumpe18}]{Lamperti17} point to a discrepancy of up to an order of magnitude between black hole masses measured from virial relations and the \mbh-\sis\ relation, in full agreement with the main conclusion of this work.

\label{lastpage}
\end{document}